\documentclass[11pt,a4paper]{article}
\pdfoutput=1
\usepackage{jheppub}
\usepackage{amsmath,amssymb}
\usepackage{slashed}
\usepackage{subfigure}
\usepackage{url}
\usepackage{bigints}
\newcommand{\beq}{\begin{equation}}
\newcommand{\eeq}{\end{equation}}

\newcommand{\be}{\begin{eqnarray}}
\newcommand{\ee}{\end{eqnarray}}
\long\def\hidestart#1\hideend{}
\allowdisplaybreaks
\title
{Physical observables from boundary artifacts: scalar glueball 
in Yang-Mills theory}
\author{Abhishek Chowdhury$^{a}$,}
\author{A. Harindranath$^{b}$ and}
\author{Jyotirmoy Maiti$^{c}$}

\affiliation{$^{a}$Department of Physics, Raja Narendra Lal Khan Women's
College,\\
Gope Palace, P.O. Vidyasagar University, Paschim Medinipur (District), 721102,
India}

\affiliation{$^{b}$Theory Division, Saha Institute of Nuclear Physics, \\
 1/AF Bidhan Nagar, Kolkata 700064, India}

\affiliation{$^{c}$Department of Physics, Barasat Government College,\\
10 KNC Road, Barasat, Kolkata 700124, India}

\emailAdd{abhi109@gmail.com}
\emailAdd{a.harindranath@saha.ac.in}
\emailAdd{jyotirmoy.maiti@gmail.com}

\date{January 28, 2016}
\abstract {By relating the functional averages of a generic scalar operator
in simulations with Open (O) and Periodic (P) boundary conditions 
(BCs) respectively for $SU(3)$ lattice gauge theory, we show that the
scalar glueball mass and the glueball to vacuum matrix element can be extracted
very efficiently from the former. Numerical results are compared with those
extracted from the two point function of the time slice energy density 
(both PBC and OBC). The scaling properties of the mass and the matrix element
are studied with the help of Wilson (gradient) flow.}

\begin{document}

\maketitle

\section{Motivation}
In confronting experimental data, lattice Quantum Chromodynamics has achieved 
remarkable progress over the years. Nevertheless certain problems emerge as 
the continuum limit is approached, a major difficulty being the spanning of 
gauge configurations over different topological sectors when periodic boundary 
condition (PBC) is used in the temporal direction. To overcome this problem,
open boundary condition (OBC) in the temporal direction has been proposed
recently \cite{open0, open1, open2}.     
In order to avoid undesirable effects in the spectrum of the Hamiltonian, 
boundary conditions are retained to be periodic for the three-dimensional 
space, which ensures 
that the transfer matrix is unaltered. Earlier, the advantage of 
OBC over PBC has been recognized and profitably
utilized in Density Matrix Renormalization Group calculations \cite{white} 
applied to condensed matter systems. Later, some undesirable features of OBC 
for systems that do not possess an energy gap have been recognized and
investigated in 
detail \cite{shibata}. However, the systems under our consideration, namely,
pure Yang-Mills theory and QCD fortuitously possess mass gaps. 
Moreover, 
OBC yields some unexpected extra dividends as we demonstrate in this work.
For example, by studying the boundary artifacts in the vacuum 
expectation value of a one point function, one can extract the mass and 
operator matrix elements which are usually extracted from a two point
function. We illustrate this idea in the context of the calculation of scalar
glueball
mass and glueball to vacuum matrix element in SU(3) Lattice Yang-Mills theory. 

For extracting the mass and the matrix element in lattice Yang-Mills 
theory, smoothing of gauge fields is essential. The Wilson (gradient)
flow \cite{wf1,wf2,wf3} provides a very convenient tool
for smoothing, with a rigorous mathematical underpinning.    
Unlike the conventional smearing techniques, the Wilson flow 
provides a common reference scale. Thus by choosing a particular flow time, 
one can study the scaling properties of observables extracted from lattice
calculations employing different lattice spacings. It is interesting to
perform such scaling studies for the glueball mass and the glueball to
vacuum matrix element. Another noise reduction technique was recently
investigated \cite{Majumdar:2014cqa} in the extraction of glueball masses. 
\section{Relation between correlation functions in OBC and PBC}
\label{sec2}
We start from the standard Wilson action for SU(3) lattice gauge theory on a 
$L^3 \times T$ lattice with periodic boundary conditions in all directions
\be
S_{\rm PBC}~=~ \frac{2}{g^2}\sum_{{ x}}&& 
\sum_{\mu< \nu} {\rm tr} \left [
1 - {\rm Re}~ U_{\mu \nu}(x) \right]
\ee
where $U_{\mu \nu} (x)$ denotes the product of the link variables around a 
plaquette $P$ in the $\mu$ -- $\nu$ plane
whose lower left hand corner is at $x$ and the sum is over all oriented 
plaquettes on the lattice.

Using transfer matrix arguments \cite{luschertm,sint,open1}, 
keeping in mind that with open boundary condition in the temporal
direction, there are no temporal
links connecting the time slice $x_0=T-1$ to the time slice $x_0=0$,
one arrives at the action for 
SU(3) lattice gauge theory with open boundary condition
\be
S_{\rm OPEN}~=~ \frac{2}{g^2}\sum_{{ x}}&&
\sum_{\mu< \nu} w(P) ~ {\rm tr} ~\left [
1 - {\rm Re}~ U_{\mu \nu}(x) \right]
\ee
where $w(P)$ is equal to 1 except for the spatial plaquette at time
$x_0=0$ and $T-1$ which have weight $\frac{1}{2}$ and here the sum runs over
the plaquettes having their corners within the time interval $[0,T-1]$.

Thus we find that $S_{\rm PBC} ~ = ~ S_{\rm OPEN} ~ + ~\Delta S$ where
\be
\Delta S = \frac{2}{g^2} \sum_{{\mathbf x}}&& \hspace{-5mm}\Bigg\{\Bigg.
\frac{1}{2}\sum_{i < j}\Big({\rm tr}
\left[1-{\rm Re}~U_{ij}({\mathbf x},T-1)\right] + {\rm tr}
\left[1-{\rm Re}~U_{ij}({\mathbf x},0)\right]\Big)\nonumber\\
&+&\sum_i {\rm tr}\left[1-{\rm Re}~U_{i4}({\mathbf x},T-1)
\right]\Bigg.\Bigg\}.
\label{deltaS}
\ee
Note that $e^{-\Delta S}$ is the transfer matrix element \cite{luschertm} 
between the
time slices $T-1$ and 0. Denoting the general transfer matrix element between
time slices $x_0$ and $x_0+1$ by $e^{-L^3 H_m(x_0)}$, we have
\be
 H_m(x_0) = \frac{2}{g^2}  \frac{1}{L^3} \sum_{{\mathbf x}}&& \hspace{-5mm}\Bigg\{\Bigg.
\frac{1}{2}\sum_{i < j} \Big({\rm tr}
\left[1-{\rm Re}~U_{ij}({\mathbf x},x_0)\right] + {\rm tr}   
\left[1-{\rm Re}~U_{ij}({\mathbf x},x_0+1)\right]\Big)\nonumber\\
&+&\sum_i {\rm tr}\left[1-{\rm Re}~U_{i4}({\mathbf x},x_0)
\right]\Bigg.\Bigg\}.
\label{deltaH}
\ee

An observable which we have extensively studied in our previous work
\cite{Chowdhury:2014kfa} is
the time slice energy density ${\overline E}(x_0)$ which was used as an
interpolating operator to calculate the scalar glueball mass. 
We adopt a symmetric definition of the time slice energy density as
\be
{\overline E}(x_0) = \frac{2}{g^2} \frac{1}{L^3} \sum_{{\mathbf x}}&&
\hspace{-5mm}\Bigg\{\Bigg.
\sum_{i< j} {\rm tr}\left[1-{\rm Re}~U_{ij}({\mathbf x},x_0)
\right]\nonumber\\
&+&\frac{1}{2}\sum_i \Big({\rm tr}\left[1-{\rm Re}~U_{i4}({\mathbf x},x_0)
\right] + {\rm tr}\left[1 - {\rm Re}~ U_{i4}({\mathbf x},x_0-1)\right]\Big)
\Bigg.\Bigg\}
\ee
where $U_{\mu\nu}({\mathbf x},x_0)$ denotes the oriented plaquette in the
$\mu-\nu$ plane with $({\mathbf x},x_0)$ at its lower-left corner.
The time slice $x_0$ is arbitrary for periodic boundary condition but it is
restricted within the bulk when open boundary condition is imposed in the
temporal direction. For the latter case, the definitions of ${\overline E}(x_0)$
with $x_0$ lying on the boundaries are given by
\be
{\overline E}(x_0=0) = \frac{2}{g^2} \frac{1}{L^3}\sum_{{\mathbf x}}\frac{1}{2}
\Bigg\{\Bigg.&&\hspace{-5mm}
\sum_{i < j} {\rm tr} \left[1-{\rm Re}~U_{ij}({\mathbf x},0)
\right]\nonumber\\ &+& 
\sum_i {\rm tr} \left[1-{\rm Re}~U_{i4}({\mathbf x},0)
\right]\Bigg\}\\{\rm and}\quad
{\overline E}(x_0=T-1) = \frac{2}{g^2} \frac{1}{L^3}\sum_{{\mathbf x}}\frac{1}{2}
\Bigg\{\Bigg.&&\hspace{-5mm}
\sum_{i < j} {\rm tr} \left[1-{\rm Re}~U_{ij}({\mathbf x},T-1)
\right]\nonumber\\ &+&
\sum_i {\rm tr}\left[1-{\rm Re}~U_{i4}({\mathbf x},T-2)
\right]\Bigg\}.
\ee

To find a relation between
$\left\langle {\cal O}(x_0)\right\rangle_{\rm OPEN}$ and 
$\left\langle{\cal O}(x_0)\right\rangle_{\rm PBC}$ where 
${\cal O}$ is a generic scalar operator 
we start from
\be
\left\langle {\cal O} \right\rangle_{\rm OPEN}
= ~
\frac{\bigintss {\mathcal D}U^\prime~{\cal O}(x_0)~e^{-S_{\rm OPEN}}} 
{\bigintss{\cal D}U^\prime  ~e^{-S_{\rm OPEN}} }
\ee
where the measure ${\cal D}U^\prime$ excludes the measures for the links
connecting the boundary time slices ($x_0=T-1$ and $x_0=0$). However, as the
integrands in both numerator and denominator do not depend on these links, we
can include their measures without altering the result. This leads us to
replace ${\cal D}U^\prime$ by ${\cal D}U$ which is the measure in case of PBC.
Thus  
\be
\left\langle {\cal O}(x_0) \right\rangle_{\rm OPEN}
&=& ~ \frac{\bigintss {\mathcal D}U~{\cal O} (x_0)~e^{-S_{\rm OPEN}}} 
{\bigintss{\cal D}U ~e^{-S_{\rm OPEN}} }\nonumber\\
&=& ~ \frac{\bigintss {\mathcal D}U~{\cal O}(x_0)~e^{-S_{\rm PBC} + \Delta S}} 
{\bigintss{\cal D}U ~e^{-S_{\rm PBC}+\Delta S} }
\ee
where the exponents of the integrands in both numerator and denominator on the
right hand side consist of fields on a periodic lattice.
Thus
\be
&&\left\langle {\cal O}(x_0) \right\rangle_{\rm OPEN}
={\bigintss {\cal D}U~  {\cal O}(x_0)~ e^{-S_{\rm PBC}+\Delta S }\Big/ 
\bigintss {\cal D}U~ e^{-S_{\rm PBC}}
\over
\bigintss {\cal D}U~ e^{-S_{\rm PBC}+\Delta S } \Big/ 
\bigintss {\cal D}U~ e^{-S_{\rm PBC}} }\nonumber\\
&=& \left\langle {\cal O}(x_0)\right\rangle_{\rm PBC} 
~+~\frac{\left\langle{\cal O}(x_0)~e^{\Delta S }  \right\rangle_{\rm PBC}^{\rm connected}}{\left\langle e^{\Delta S}
\right\rangle_{\rm PBC}}~\label{mastere} \\
 &=& \left\langle{\cal O} (x_0)\right\rangle_{\rm PBC} ~+~ \frac{1}{r}
\left\langle {\cal O}(x_0)~e^{L^3 H_m(T-1)}\right\rangle_{\rm PBC}^{\rm connected}
 \label{oneptwop}
\ee
where $r= \langle e^{\Delta S}\rangle_{\rm PBC} = \langle e^{L^3 H_m(T-1)}
\rangle_{\rm PBC}$.

As $e^{L^3 H_m(x_0)}$ is also a scalar operator, from eq. \ref{oneptwop} we have
\be
\left\langle {\cal O}(x_0)\right\rangle_{\rm OPEN} 
\approx  \left\langle {\cal O}(x_0)\right\rangle_{\rm PBC}
~+~2C_{1}^{\prime} ~e^{-m T/2}~\cosh m\Big(\frac{T}{2}-1-x_0\Big).
\ee
where $m$ is the scalar glueball mass.

In comparison, the two point function for the time slice energy density in
the case of PBC behaves as
\be 
 \left\langle{\overline E}(x_0) {\overline E}(x_0=0)  \right\rangle_{\rm
PBC}
~\approx~   C_0~+~ 2 C_1 ~ e^{-m T/2}~\cosh m\Big(\frac{T}{2}-x_0\Big).
\ee 
where
\be
C_1 = \frac{\left|\langle
0\left|{E}\left(0\right)\right|G\rangle\right|^2}{2m} = 
\frac{C^2}{2m}.
\ee

Thus we find that one can extract certain two-point correlators by analyzing
the data for the functional average of a scalar operator computed with open
boundary (in the temporal direction) in the region of $x_0$ where it differs,
due to the breaking of translational invariance, from the same computed with
periodic boundary. Same technique can, in principle, be used to compute any
n-point correlator in the scalar channel. In the case of lattice 
QCD with OBC, same technique can also be used to extract the mass of the 
lowest two-pion state. A recent simulation with 2+1 
flavors \cite{Bruno:2014jqa} however 
encountered large scaling violations which unfortunately made such an 
extraction not possible.

We note that while the extraction of the glueball mass from the one point
function is as straightforward as from the two point function, the
extraction of the glueball to vacuum matrix element from the former is not
as straightforward as from the latter. However for the operators 
($\overline{E}$ and $H_m$) used in this work, this becomes possible in
the region of very small lattice spacing. See the appendix for the discussion.
This is supported by our numerical results.    
   
\section{Determination of the mass and the matrix element}
In this section, we discuss the methods for extraction of the glueball mass and
the glueball to vacuum matrix element from the two-point (PBC and OBC) and the
one point (OBC) correlation functions. 
\subsection{Periodic Boundary Condition (PBC)}
In the case of PBC, the mass and the matrix element have to be extracted
from the two point correlator. Because of the periodicity of the lattice we
have
\be
 \left\langle{\overline E}(x_0) {\overline E}(x_0=0)  \right\rangle_{\rm PBC}
~=~ G(x_0) ~\simeq~  C_0~+~ C_1 ~ \Big [ 
e^{-mx_0}~ + ~ e^{-m(T-x_0)}
\Big ]~.
\ee  
The effective mass is calculated by solving the equation $F(m)=0$
\be
F~=~ (r_1-1)~\Big [ \cosh m(dt-1) ~ - ~   \cosh mdt \Big ]
~+~ (1-r_2)~ \Big [ \cosh m(dt+1) ~ - ~   \cosh mdt \Big ] \nonumber\\
\ee
where 
\be
r_1  ~=~ \frac{G(x_0-1)}{G(x_0)}, ~~~~ 
r_2  ~=~ \frac{G(x_0+1)}{G(x_0)}~~~~~ {\rm and}~~ dt=T/2 - x_0~.   
\ee
The coefficient has been extracted as
\be
C_1~ = ~ \frac{1}{2} \left (C_1^{(1)} ~+~ C_1^{(2)} \right )
\ee
where
\be
 C_1^{(1)}~&=&~ \frac{G(x_0) ~ - ~ G(x_0+1)}{e^{-mx_0}~+~e^{-m(T-x_0)}~
-~e^{-m(x_0+1)}~-~e^{-m(T-x_0-1)}}\\
{\rm and}\quad
 C_1^{(2)}~&=&~ \frac{G(x_0-1) ~ - ~ G(x_0)}{e^{-m(x_0-1)}~+~e^{-m(T-x_0+1)}~
-~e^{-m(x_0)}~-~e^{-m(T-x_0)}}.
\ee

\subsection{Open Boundary Condition (OBC)}

In the case of OBC, the mass and the matrix element can be extracted
from the one point correlator as well as from the two point function as
discussed in the previous section. In the case of two point correlator, 
because of the lack of translational invariance, we have
\be
 \left\langle{\overline E}(x_0+x_0^\prime) {\overline E}(x_0^\prime)  \right\rangle_{\rm
OBC}
~=~ G(x_0) ~\simeq~  C_0~+~ C_1 ~  
e^{-mx_0}~
\ee
where $x_0^\prime$ is well within the bulk.  
Thus the effective mass is given by
\be
m~=~ {\rm ln}~ \frac{G(x_0-1) - G(x_0)}{G(x_0) - G(x_0+1)}.
\ee
We extract the coefficient as
\be
C_1~ = ~ \frac{1}{2} \left (C_1^{(1)} ~+~ C_1^{(2)} \right )
\ee
where
\be
 C_1^{(1)}~=~ \frac{G(x_0) ~ - ~ G(x_0+1)}{e^{-mx_0}~
-~e^{-m(x_0+1)}}\quad
{\rm and}\quad 
 C_1^{(2)}~=~ \frac{G(x_0-1) ~ - ~ G(x_0)}{e^{-m(x_0-1)}~
~-~e^{-mx_0}}.
\ee

\section{Numerical results}
\begin{table}[h]
\centering{
\begin{tabular}{|c|l|l|l|l|l|l|}
\hline \hline
Lattice & Volume & $\beta$ & $N_{\rm cnfg}$  &$\tau$&$a[{\rm fm}]$ 
& $t_{0}/a^2$\\
\hline\hline
{$O_1$}&{$24^3\times48$}&{6.21} &{3970}  &{3}& {0.0667(5)} &
{6.207(15)}\\
\hline
{$O_2$}&{$32^3\times64$}&{6.42} &{3028} &{4}&{0.0500(4)} &
{11.228(31)}\\
\hline
{$O_3$}&{$48^3\times96$}&{6.59} &{2333} &{5}&{0.0402(3)} &
{17.630(53)}\\
\hline
{$O_4$}&{$64^3\times128$}&{6.71} &{181} &{10}&{0.0345(4)} &
{24.279(227)}\\
\hline
{$P_1$}&{$24^3\times48$}&{6.21} &{3500} &{3}&{0.0667(5)} &
{6.197(15)}\\
\hline
{$P_2$}&{$32^3\times64$}&{6.42} &{1958} &{4}&{0.0500(4)} &
{11.270(38)}\\
\hline
{$P_3$}&{$48^3\times96$}&{6.59} &{295} &{5}&{0.0402(3)} &
{18.048(152)}\\
\hline\hline
\end{tabular}
\caption{$N_{\rm cnfg}$ is 
the number of configurations, $\tau$ is the trajectory length used in the HMC
simulation and $t_{0}/a^2$ is the dimensionless reference Wilson flow time.
$O$ and $P$ refer to ensembles with open and periodic boundary conditions in
the temporal direction.}
\label{table1}
}
\end{table}

SU(3) gauge configurations in lattice Yang-Mills theory are generated
with open boundary condition (denoted by O) at different lattice volumes (by
lattice volume we mean total number of lattice points) and gauge
couplings using the openQCD program \cite{openqcd}. For
comparison purposes, by implementing periodic boundary condition
in temporal direction in the openQCD package, we have also generated gauge 
configurations (denoted by P) for several of the same lattice parameters.
Simulation details are given in table \ref{table1}. The parameter $t_0$
is defined in the context of Wilson flow \cite{wf1,wf2,wf3} which is used to
smooth the gauge configurations. The implicit equation
\beq
\big\{t^2\langle\overline{E}(T/2)\rangle\big\}_{t=t_0} = 0.3
\eeq 
with $t$ and $T$ being respectively the Wilson flow time and
the temporal extent of the lattice, defines a reference flow time $t_0$
which provides a reference scale to extract physical quantities from
lattice calculations.
The effectiveness of the Wilson flow in the extraction of topological
susceptibility \cite{opentopo}, glueball mass \cite{Chowdhury:2014kfa}
and topological charge density correlator \cite{Chowdhury:2014mra}
has been demonstrated recently.

\begin{figure}[h]
\centering   
\includegraphics[width=4in,clip]{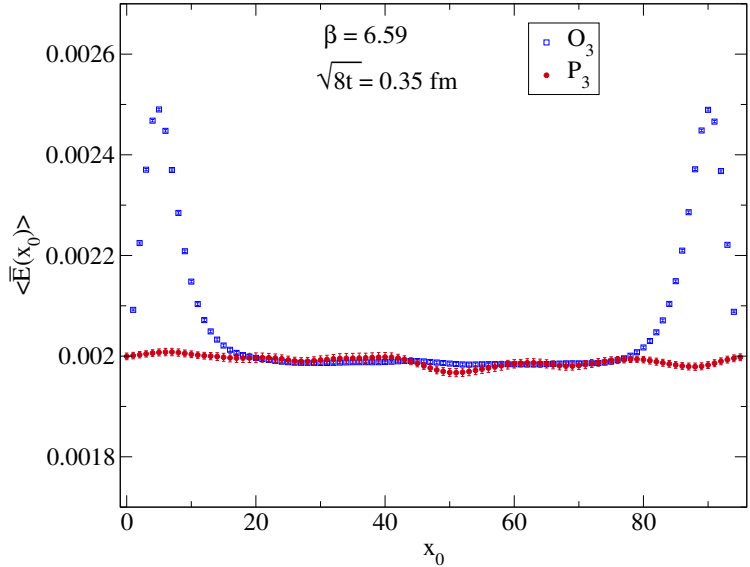}
\caption{Plot of $\langle\overline{E}(x_0)\rangle$ versus $x_0$ at flow 
time $t=t_0$ at $\beta=
6.59$ and lattice volume $48^3\times 96$ for ensemble $O_3$ (filled circle)    
and ensemble $P_3$ (filled square).}
\label{edtime}  
\end{figure}
In fig. \ref{edtime} we plot the configuration average of the time slice
energy density averaged over the spatial volume
$\langle\overline{E}(x_0)\rangle$ versus the time slice $x_0$
at the reference flow
time $t=t_0$ \cite{wf2} at $\beta=
6.59$ and lattice volume $48^3\times 96$ for ensemble $O_3$ (filled circle)
and ensemble $P_3$ (filled square). 
We find
that in the boundary region at $x_0=0$ ,
$\langle\overline{E}(x_0)\rangle_{\rm OBC}$ rises above 
and then decays to
$\langle\overline{E}(x_0)\rangle_{\rm PBC}$. Similar behaviour is found in
the boundary region $x_0=T-1$.  
As explained in sec. \ref{sec2}, the decay rate is determined by the mass of the
lowest excitation in the scalar channel, namely the glueball mass.

\begin{figure}[h]
\begin{minipage}[h]{0.45\linewidth}
\hspace{-.5cm}
\includegraphics[width=3.25in,clip]
{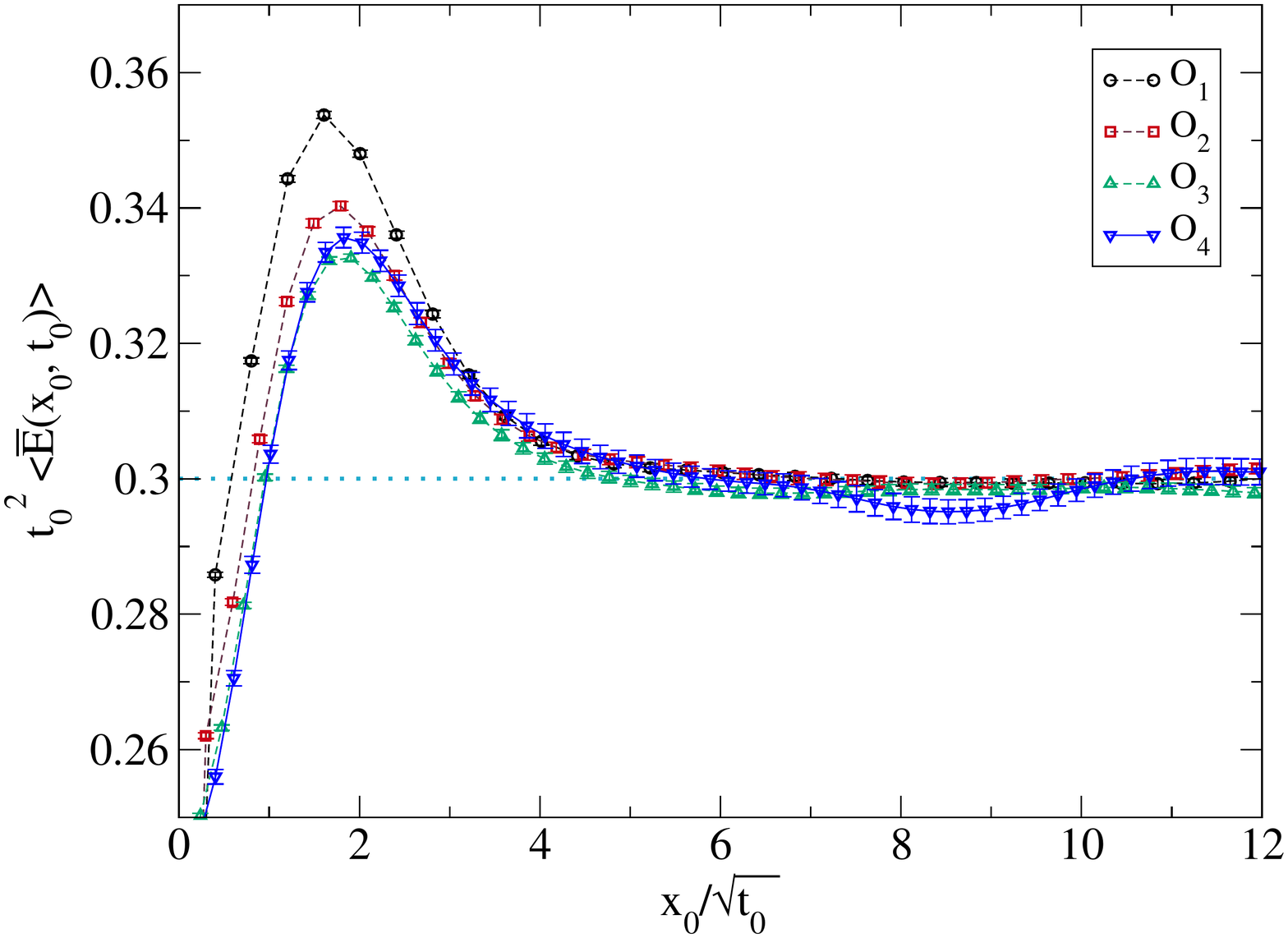}
\end{minipage}
\begin{minipage}[h]{0.45\linewidth}
\hspace{.5cm}
\includegraphics[width=3.25in,clip]
{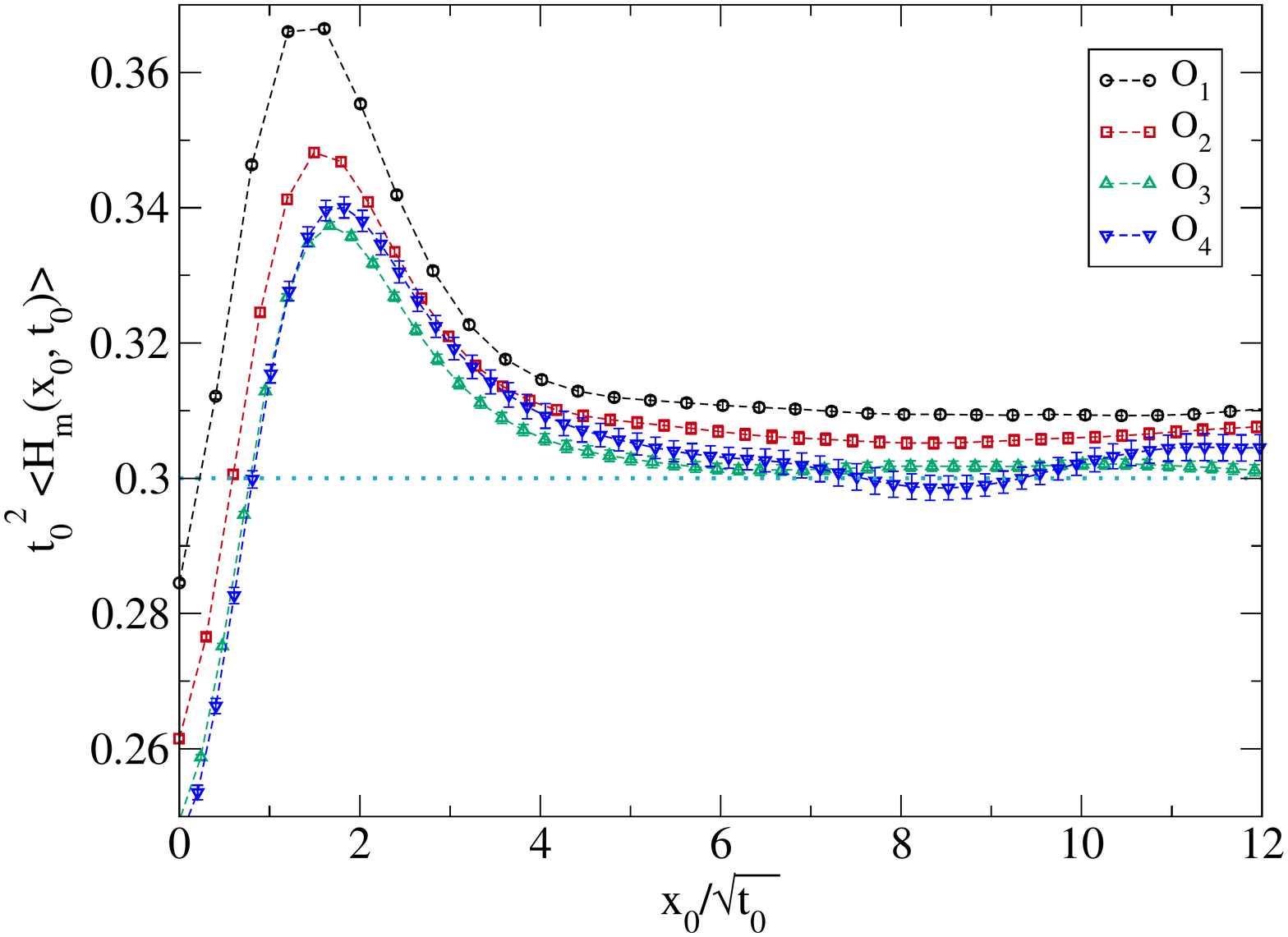}
\end{minipage}
\caption{Plot of $t_0^2 \langle\overline{E}(x_0, t_0)\rangle$ 
(left) and $t_0^2 \langle{H_m}(x_0, t_0)\rangle$
(right)
versus
$x_0/\sqrt{t_0}$ for the ensembles $O_1$, $O_2$, $O_3$ and $O_4$.} 
\label{tsqehdtime}
\end{figure}
In order to reliably determine the glueball mass from the decay of 
$\langle\overline{E}(x_0)\rangle_{\rm OBC}$ and 
$\langle{H_m}(x_0)\rangle_{\rm OBC}$
first we need to verify their
scaling behaviour. Towards this goal, 
in fig. \ref{tsqehdtime} (left) we plot $t_0^2 \langle\overline{E}(x_0, t_0)\rangle$
versus $x_0/\sqrt{t_0}$ for the
ensembles $O_1$, $O_2$, $O_3$ and $O_4$. We note that, except for the
largest
lattice spacing, the data exhibit 
excellent scaling behaviour in the tail region from where one can extract
the glueball mass. In comparison, in 
fig. \ref{tsqehdtime} (right), the scaling behaviour of $t_0^2
\langle{H_m}(x_0, t_0)\rangle$ is shown. 
The difference between these observables decreases as the lattice spacing
decreases.
The slightly worsened scaling
behaviour of the latter
for the two relatively larger lattice spacings can be readily
attributed to the following. The expression for $H_m$ follows directly from
the lattice action used in this work, namely, the unimproved Wilson
gauge action. On the other hand, the expression for ${\overline E}$ uses the
clover definition of the lattice field tensor. Thus the difference in
scaling behaviour exhibited in the left and right parts of 
fig. \ref{tsqehdtime} results from the use of an improved versus an unimproved
operator, which diminishes as lattice spacing goes to zero. The reason for the
non-smooth behaviours of both $\overline{E}$ and $H_m$ in case of the ensemble
$O_4$ is due to the lack of statistics.   

\begin{figure}[h]
\centering   
\includegraphics[width=4in,clip] 
{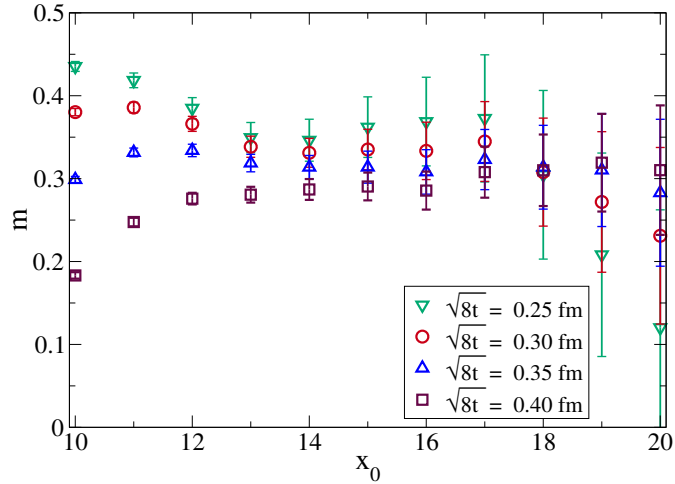}
\caption{The scalar Glueball effective mass $m(0^{++})$ 
as a function of $x_0$ extracted from the one-point function of $\overline{E}$
for four values of Wilson flow times for the ensemble $O_3$.}
\label{effmass}
\end{figure}\vskip2mm

In fig. \ref{effmass} we show an example of the effective glueball mass
determination from the one-point function. The effective mass $m(0^{++})$ is plotted versus $x_0$ for four different  Wilson flow times for the
ensemble $O_3$. We find that there is an optimum window of Wilson flow
time within which the glueball mass can be reliably extracted from the 
effective mass plot. For lower values of Wilson flow time, the smearing is 
not able to remove unwanted contributions completely and the plateau region 
is too narrow. For higher
values of Wilson flow time also plateau is too narrow, presumably due to
over smearing resulting in the overlap of the two glueball interpolation
operators in the correlation function. Ideally the plateau region should be 
independent of the flow
time chosen, but we observe that an overlapping (within the statistical
errors) and extended plateau exists only in the region of flow time between 
0.3 fm and 0.35 fm. In the following we will present the glueball mass and
glueball to vacuum matrix element extracted for these two values of the flow
time. 

\begin{table}
\begin{tabular}{|l|l|l|l|l|l|}
\hline \hline
Lattice & Correlator &\multicolumn{2}{c|}{Mass $m_G$ (MeV)} & 
\multicolumn{2}{c|}{Coefficient
$r_0^3C$}\\
\cline{3-6}
&  & {$\sqrt{8t}$=0.3fm} & {$\sqrt{8t}$=0.35fm} & {$\sqrt{8t}$=0.3fm} &
{$\sqrt{8t}$=0.35fm}\\
\hline
{$O_{1}$}& One point ($H_m$)&{1729 (67)} &{1649 (56)} &{98 (8)} &{85 (6)} \\
\cline{2-6}
& One point (${\overline E}$)&{1690 (94)} &{1650 (80)} &{104 (14)} &{95 (11)} \\
\cline{2-6}
& Two point &{1626 (186)} &{1501 (96)} &{103 (17)} &{88 (7)} \\
\hline
{$P_{1}$}& Two point &{1625 (92)} &{1594 (72)} &{106 (10)} &{100 (7)} \\
\hline
{$O_{2}$}&  One point ($H_m$)&{ 1700 (70)} &{1645 (59)} &{102 (8)} &{90 (6)} \\
\cline{2-6}
&One point (${\overline E}$)&{1710 (84)} &{1629 (48)} &{113 (12)} &{97 (5)} \\
\cline{2-6}
&  Two point&{1587 (234)} &{1458 (115)} &{101 (21)} &{85 (8)} \\
\hline
{$P_{2}$}& Two point&{1552 (61) } &{1506 (71)} &{99 (5)} &{91 (6)} \\
\hline
{$O_{3}$} & One point($H_m$)&{1640 (101)} &{1551 (81)} &{103 (14)} &{88 (9)} \\
\cline{2-6}
& One point (${\overline E}$)&{1625 (85)} &{1540 (69)} &{108 (12)} &{93 (9)} \\
\cline{2-6}
&  Two point&{1616 (254)} &{1465 (161)} &{99 (20)} &{84 (12)} \\
\hline
{$P_{3}$} & Two point&{1467 (181)} &{1421 (182)} &{94 (16)} &{85 (16)} \\
\hline
{$O_{4}$} & One point ($H_m$)&{1818 (87)} &{1752 (177)} &{140 (13)} &{125
(27)} \\
\cline{2-6}
& One point (${\overline E}$)&{1783 (141)} &{1711 (273)} &{144 (24)} &{126 (48)} \\
\cline{2-6}
&  Two point&{1521 (513)} &{1459 (488)} &{90 (35)} &{83 (33)} \\
\hline\hline
\end{tabular}
\caption{The lowest scalar glueball mass ($m_G$) in MeV and the glueball
to vacuum matrix element in unit of the Sommer parameter $r_{0}$ 
extracted from correlators at Wilson flow times 
$\sqrt{8t}$ = 0.3 fm and 0.35 fm. One point and Two point 
refer to the one point and the two point correlator 
of ${\overline E}$. }\label{t2}
\end{table}

In table \ref{t2} we present the values of the lowest scalar glueball 
mass $m_G~(=m/a)$
in MeV and the glueball
to vacuum matrix element in unit of the Sommer parameter $r_{0}$ 
extracted from correlators at Wilson flow times 
$\sqrt{8t}$ = 0.3 fm and 0.35 fm. One point 
refers to the one point function  of $H_m$ and ${\overline E}$ and Two point refers to the two point correlator 
of ${\overline E}$. 
Wilson (gradient) flow is known \cite{wf1, wf2, wf3} to have the consequence 
that the expectation
values of local gauge invariant operators constructed from the gauge field
at
positive flow time are ultraviolet finite. Thus we expect the
glueball mass and the glueball to vacuum matrix element extracted at a fixed 
flow time at
different lattice spacings to exhibit scaling (provided lattice artifacts
are negligible). The glueball mass should be 
independent of the flow time but the glueball to vacuum matrix element is
expected to depend on the flow time (energy scale). The results presented in 
table \ref{t2} are consistent with these expectations within the statistical
errors and the limited range of flow times probed.  
The investigation of the relationship between the
extracted glueball to vacuum matrix element at a given Wilson flow time 
and its continuum counterpart involves a detailed numerical 
study of the behaviour of
the glueball matrix element of the energy momentum tensor and the trace anomaly
under Wilson 
flow (for related theoretical work, see for 
example, Refs. \cite{Suzuki:2013gza} and \cite{DelDebbio:2013zaa}) and 
is beyond the scope of the present work.

\begin{figure}[h]
\begin{minipage}[h]{0.45\linewidth}
\hspace{-1.5cm}
\includegraphics[width=4in,clip]
{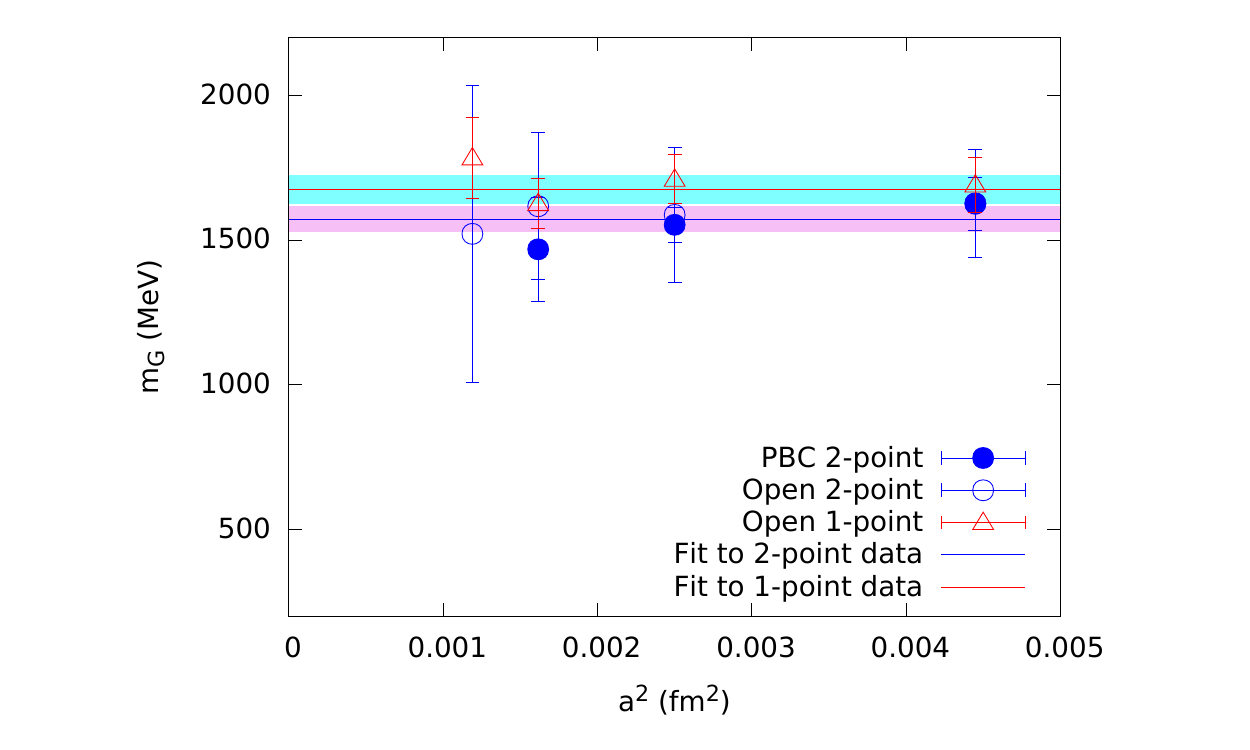}
\end{minipage}
\begin{minipage}[h]{0.45\linewidth}
\hspace{-0.5cm}
\includegraphics[width=4in,clip] 
{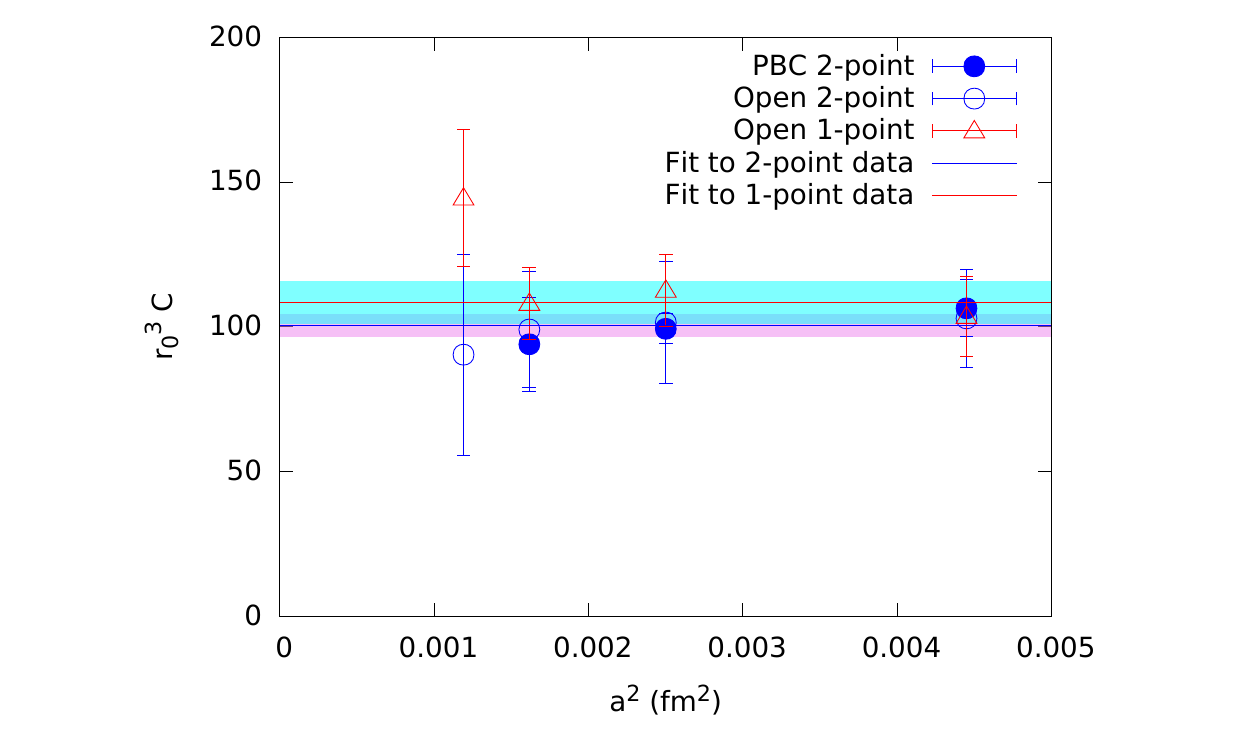}
\end{minipage}
\caption{Plot of lowest glueball mass $m_G(0^{++})$ in MeV (left) and
glueball to vacuum matrix element in unit of $r_0$ (right)
versus $a^2$ extracted from one-point (Open) and two-point correlators (Open
and PBC) of $\overline{E}$ for different lattice spacings and lattice volumes
at Wilson flow time $\sqrt{8t}=0.3$ fm. Light blue and light violet shaded
regions correspond to the error bands around the fit curves to one-point and
two-point data respectively.}
\label{masscoeffdata3}
\end{figure}

\begin{figure}[h]
\begin{minipage}[h]{0.45\linewidth}
\hspace{-1.5cm}
\includegraphics[width=4in,clip] 
{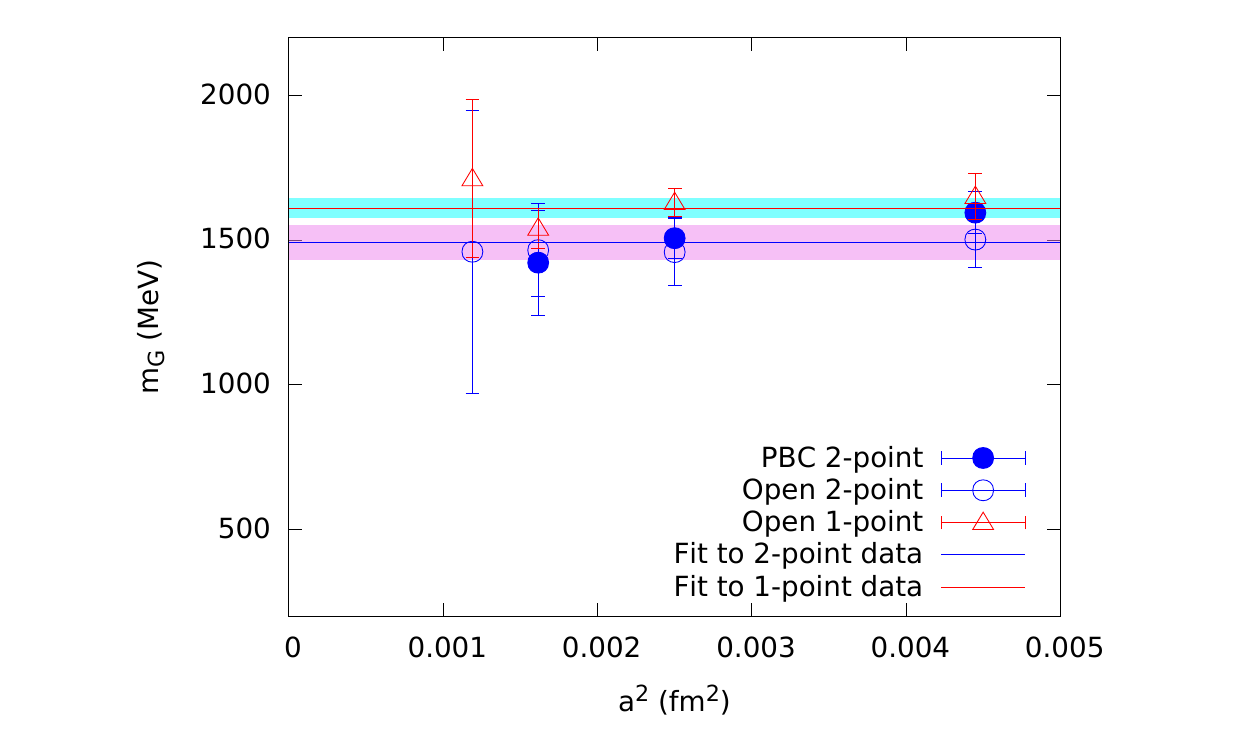}
\end{minipage}
\begin{minipage}[h]{0.45\linewidth}
\hspace{-0.5cm}
\includegraphics[width=4in,clip] 
{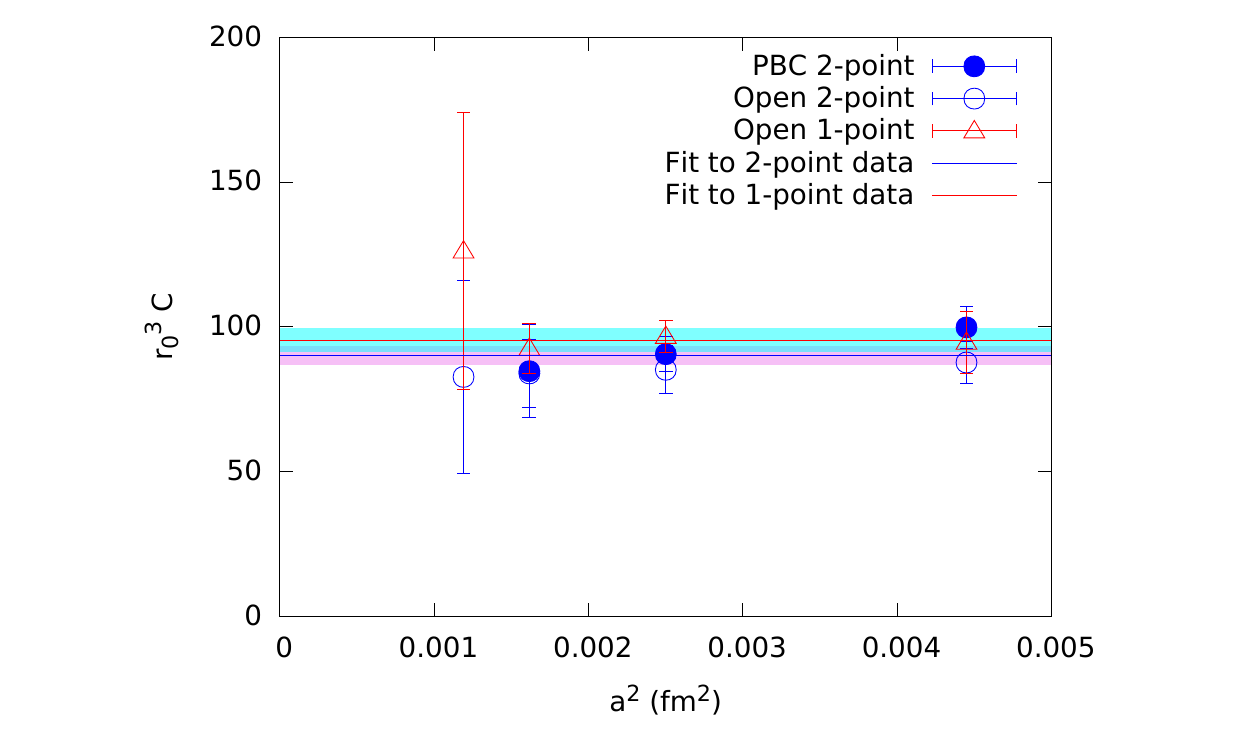}
\end{minipage}
\caption{Plot of lowest glueball mass $m_G(0^{++})$ in MeV (left) and glueball to
vacuum matrix element in unit of $r_0$  (right) versus
$a^2$ extracted from one-point (Open) and two-point correlators (Open and PBC)
of $\overline{E}$ for different lattice spacings and lattice volumes at Wilson
flow time $\sqrt{8t}=0.35$ fm. Light blue and light violet shaded regions
correspond to the error bands around the fit curves to one-point and two-point
data respectively.}
\label{masscoeffdata35}
\end{figure}

In fig. \ref{masscoeffdata3}, we have presented the variation, with $a^2$, of
lowest glueball mass and glueball to vacuum matrix element in unit of Sommer
parameter $r_0$ extracted from one-point (Open) and two-point correlators 
(Open and PBC) of $\overline{E}$ for different lattice spacings and lattice
volumes at Wilson flow time $\sqrt{8t}=0.3$ fm. Also shown are separate fits to
one-point and two-point data sets. Light blue and light violet shaded regions
correspond to the error bands around the fit curves to one-point and two-point
data respectively. Corresponding data at Wilson flow time $\sqrt{8t}=0.35$ fm
are plotted in fig. \ref{masscoeffdata35}.

Note that with PBC, glueball observables can be extracted 
only from the two point correlator. With PBC, the signal can be extracted 
after performing source averaging and hence the statistical error can be 
reduced. However, when one simulates at lower lattice spacings it becomes 
increasingly difficult to generate statistically independent configurations 
with PBC in the temporal direction and hence the statistical accuracy begins 
to suffer. This problem is overcome with OBC in the temporal directions. 
However in this case translational invariance is lost close to the boundary 
and hence one can perform source averaging only over the spatial volume and 
a few temporal slices well inside the bulk. This effectively increases
the statistical error. With two point correlator, the disconnected contribution
has to be subtracted numerically and hence this contributes to the 
increase of statistical errors since one is dealing with the subtraction 
involving two large quantities.

We have shown that, with OBC, the glueball observables can be extracted
alternatively from the one point correlation function as well. 
The automatic subtraction of the disconnected part 
in the case of one point function leads to smaller statistical 
error compared to the extraction
from two point correlation function with same boundary condition as exhibited
in table \ref{t2} and figures \ref{masscoeffdata3} and \ref{masscoeffdata35}.
One should, of course, keep in mind that this method for the
calculation of n-point correlators involving the energy density is
applicable only in the scalar channel.  

\begin{table}
\centering
\begin{tabular}{|l|l|l|l|l|}
\hline \hline
 Correlator &\multicolumn{2}{c|}{Mass $m_G$ (MeV)} & 
\multicolumn{2}{c|}{Coefficient
$r_0^3C$}\\
\cline{2-5}
  & {$\sqrt{8t}$=0.3fm} & {$\sqrt{8t}$=0.35fm} & {$\sqrt{8t}$=0.3fm} &
{$\sqrt{8t}$=0.35fm}\\
\hline
 One point (${H_m}$) & {1701 (44)} & {1628 (36)} & {100 (5)} & {88 (4)} \\
\hline
One point (${\overline E}$) & {1674 (51)} & {1610(35)} & {108 (7)} & {95 (4)} \\
\hline
 Two point & {1572 (46)} &{1490 (61)} &{100 (4)} & {90 (3)} \\
\hline\hline
\end{tabular}
\caption{Fit results for glueball mass and glueball to vacuum matrix
element.}\label{t3}
\end{table}

It is noticeable that the results (for both mass and matrix element separately)
extracted from the two point correlation functions with periodic and open
boundary conditions are very close to each other. However, there is
systematically an upward shift in the results obtained from one point
correlation functions (doable only with open boundary condition), although not
beyond the statistical errors in most cases. We have treated the data sets
extracted from two point and one point correlation functions separately in the
fitting procedure. It is also noteworthy that for the range of lattice spacings
explored in this work, scaling violations are within the statistical
uncertainty of our data. This led us to perform just constant fits to the data.
While fitting the data (both mass and matrix element) for one point correlator,
we have excluded $\beta=6.71$ as it somewhat deviates from the general trend.
However, we have checked that its inclusion in fitting procedure does not
change the results significantly because of large errors on the data at this
coupling. The fit results are presented in table \ref{t3}.

In our previous work on glueball mass extraction \cite{Chowdhury:2014kfa}, a
systematic study
of the variation of mass with Wilson flow time was not performed. For each
$\beta$, flow time yielding the most stable plateau was picked up. However, for
the study of glueball to vacuum matrix element it is mandatory to choose a
common flow time for all the lattice spacings. This path has been followed in
the present work. This causes the difference in results for glueball mass
obtained via the two point correlation functions in this work from that
quoted in \cite{Chowdhury:2014kfa}. However, it is gratifying to note that the
average of the glueball masses obtained with the two flow times is very close 
to the
value quoted in \cite{Chowdhury:2014kfa} and also agrees well with the value
given in \cite{meyer}. Although our results for glueball mass extracted from
one point correlation function are comparatively higher, they fall between
the results of \cite{meyer} and \cite{chen} both evaluated using two point
correlators.

\vskip .1in
{\bf Conclusions}
\vskip .1in
In lattice QCD, OBC in the temporal direction has been proposed to overcome
the difficulty in the spanning of gauge configurations over different
topological sectors. However, the lack of translational invariance in this
case can cause some inconveniences in the measurement of observables due to
boundary effects when compared to the case of PBC. In this work, we have
demonstrated that the same boundary artifacts can be exploited to yield
certain observables with greater efficiency. This is achieved by relating
the functional average of a generic scalar operator measured in the case
of OBC to that with PBC. The scalar glueball mass and the glueball to vacuum
matrix elements obtained from this observable in the case of OBC are compared
with the values extracted from the measurement of two-point function of the
time slice energy density in the case of both PBC and OBC. The Wilson
(gradient) flow is used to exhibit the scaling properties of both the mass
and the matrix element.     

\vskip .1in
{\bf Acknowledgements}
\vskip .1in
Cray XT5 
and Cray XE6 systems supported by the 11th-12th Five Year
Plan Projects of the Theory Division, SINP under the Department of 
Atomic Energy, Govt. of India, are used to perform  all the numerical 
calculations reported in this work.
We thank Richard Chang for the prompt maintenance of the systems and the help 
in data management.
We are deeply indebted to Martin L\"{u}scher for the suggestion that glueball 
mass can be extracted 
from the expectation value of the energy density in Yang-Mills theory with OBC.
This work was in part based on the publicly available lattice gauge theory 
code {\tt openQCD} \cite{openqcd}.

\enlargethispage{\baselineskip}

\appendix
\section{Connection between matrix elements}
In the following, we show that, with a set of approximations, 
one can understand why the glueball to vacuum matrix element of 
both $H_m$ and ${\overline E}$ calculated from the one point functions agree 
(within statistical errors) with that calculated from the two point function
of ${\overline E}$.   

Consider the two point correlator appearing in eq. \ref{oneptwop} for the
one point function of ${\overline E}(x_0)$.
In this correlator, for $x_0$, in ${\overline E}(x_0)$, far away from the 
boundaries the last
term in the expression for $\Delta S$ given in eq. \ref{deltaS} can be
approximated as
\be
\frac{1}{g^2} ~ \sum_{{\mathbf x}}&&\hspace{-2mm}\sum_{i}{\rm tr}
\left[1-{\rm Re}~U_{i4}({\mathbf x},T-1)\right]\nonumber\\
\approx\frac{1}{g^2} ~ \sum_{{\mathbf x}}&&\hspace{-2mm}\Bigg\{\Bigg.
\frac{1}{4}\Big(\sum_{i}{\rm tr}\left[1-{\rm Re}~U_{i4}({\mathbf x},0)
\right]+\sum_i {\rm tr}\left[1-{\rm Re}~U_{i4}({\mathbf x},T-1)\right]
\Big)\nonumber\\
&&+\frac{1}{4}\Big(\sum_{i}{\rm tr}\left[1-{\rm Re}~U_{i4}({\mathbf x},T-1)
\right]+\sum_{i}{\rm tr}\left[1-{\rm Re}~U_{i4}({\mathbf x},T-2)\right]
\Big)\Bigg.\Bigg\}.\nonumber
\ee
With this approximation we find,
$\Delta S \approx \frac{L^3}{2}
\Big({\overline E}(0) ~+~{\overline E}(T-1) \Big)$. This leads us to
write
\be
&&\left\langle {\overline E}(x_0) \right\rangle_{\rm OPEN}
= \left\langle{\overline E}(x_0)\right\rangle_{\rm PBC} 
~+~\frac{\left\langle{\overline E}(x_0)~e^{\frac{L^3}{2}~
\Big({\overline E}(0) ~+~{\overline E}(T-1) 
\Big) }  \right\rangle_{\rm PBC}^{\rm connected}}
{\left\langle e^{\frac{L^3}{2}~
\Big({\overline E}(0) ~+~{\overline E}(T-1) 
\Big)}
\right\rangle_{\rm PBC}}.~\label{pbcconn}
\ee
Now in continuum limit $\overline{E}(x_0) = a\overline{E}_{YM}(x_0)$
where $\overline{E}_{YM}(x_0) = \frac{a^3}{2L^3}\sum\limits_{\mathbf x}{\rm tr}
\left\{G_{\mu\nu}\left(x\right)G_{\mu\nu}\left(x\right)\right\}$. This enables us
to write 
\be
e^{\frac{L^3}{2}~\Big({\overline E}(0) ~+~{\overline E}(T-1) \Big)} = 1 + a\frac{L^3}{2}
\Big({\overline E}_{YM}(0) ~+~{\overline E}_{YM}(T-1) \Big)+{\mathcal{O}}
\left(a^2\right).
\ee
So eq. \ref{pbcconn} can be approximated as
\be
&&\left\langle {\overline E}(x_0) \right\rangle_{\rm OPEN}\nonumber\\
&=& \left\langle{\overline E}(x_0)\right\rangle_{\rm PBC} + a^2
\frac{L^3}{2}~ \left\langle{\overline E}_{YM}(x_0)
\Big({\overline E}_{YM}(0) ~+~{\overline E}_{YM}(T-1) 
\Big)   \right\rangle_{\rm PBC}^{\rm connected}~+~{\mathcal{O}}
\left(a^3\right) \nonumber\\
&=&  \left\langle{\overline E}(x_0)\right\rangle_{\rm PBC}
~+~C_1^\prime ~e^{-mT/2}~\left\{\cosh m\Big(\frac{T}{2}-x_0\Big)~+~\cosh
m\Big(\frac{T}{2}-\left(x_0+1\right)\Big)\right\}+~\dots\nonumber\\
&\approx&  \left\langle{\overline E}(x_0)\right\rangle_{\rm PBC}
~+~2 C_1^\prime ~e^{-mT/2}~\cosh m\Big(\frac{T}{2}-x_0\Big)~+~\dots .
\label{capprox}  
\ee

As evident from fig. \ref{tsqehdtime}, $H_m(x_0)$ can be approximated by
$\overline{E}(x_0)$ in the limit of very small lattice spacing. Therefore, in
this limit one point function of $H_m$ takes the form
\be
&&\left\langle H_m(x_0) \right\rangle_{\rm OPEN}
= \left\langle H_m(x_0)\right\rangle_{\rm PBC} 
~+~\frac{\left\langle H_m(x_0)~e^{L^3 H_m(T-1) } \right\rangle_{\rm PBC}^{\rm connected}}
{\left\langle e^{H_m(T-1)}
\right\rangle_{\rm PBC}}~\nonumber \\
&&\approx \left\langle{\overline E}(x_0)\right\rangle_{\rm PBC} 
~+~\frac{\left\langle{\overline E}(x_0)~e^{L^3~{\overline E}(T-1) }
\right\rangle_{\rm PBC}^{\rm connected}}
{\left\langle e^{L^3~{\overline E}(T-1)}\right\rangle_{\rm PBC}}.
\ee
Discussions subsequent to eq. \ref{pbcconn} upto eq. \ref{capprox} follow
thereafter. 


\end{document}